\documentclass[11pt]{article}
\usepackage{graphicx}
\usepackage{cite}

\newcommand{\BABARPubYear}    {06}

\newcommand{\BABARConfNumber} {032}
\newcommand{\SLACPubNumber} {11994}

\input babarsym

\def\emu {\ensuremath{e\mu}\xspace}
\def\modepill {\ensuremath{B\to \pi\ellell}\xspace}

\def\modepiem {\ensuremath{B\to \pi\emu}\xspace}
\def\modepichll {\ensuremath{B^+\rightarrow \pi^+ \ellell}\xspace}
\def\modekll {\ensuremath{B^+\rightarrow K^+ \ellell}\xspace}
\def\modepichee {\ensuremath{B^+\rightarrow \pi^+ \epem}\xspace}
\def\modepichmm {\ensuremath{B^+\rightarrow \pi^+ \mumu}\xspace}
\def\modepichem {\ensuremath{B^+\rightarrow \pi^+ \emu}\xspace}
\def\modepizee {\ensuremath{B^0\rightarrow \pi^0 \epem}\xspace}
\def\modepizmm {\ensuremath{B^0\rightarrow \pi^0 \mumu}\xspace}
\def\modepizll {\ensuremath{B^0\rightarrow \pi^0 \ellell}\xspace}
\def\modepizem {\ensuremath{B^0\rightarrow \pi^0 \emu}\xspace}

\long\def\inst#1{\par\nobreak\kern 4pt\nobreak
    {\it #1}\par\vskip 10pt plus 3pt minus 3pt}

\begin{document}
{\pagestyle{empty}

\begin{flushright}
\babar-CONF-\BABARPubYear/\BABARConfNumber \\
SLAC-PUB-\SLACPubNumber \\
\end{flushright}

\par\vskip 5cm

\begin{center}
  \Large \bf  Search for the rare decay 
${ \boldmath B \to \pi \; \ell^+ \ell^-}$
\end{center}
\bigskip

\begin{center}
\large The \babar\ Collaboration\\
\mbox{ }\\
\today
\end{center}
\bigskip \bigskip

\begin{center}
\large \bf Abstract
\end{center}

We present results of a search for the rare flavor-changing
neutral-current decay \modepill, based on a data sample corresponding to 
209 \invfb of integrated luminosity collected with the \babar\, detector at the PEP-II \B
Factory.  We reconstruct the four exclusive \B decay modes
\modepichll and \modepizll, where $\ell$ is either an $e$ or $\mu$.
We find no evidence for a signal, and we obtain the upper limit 
at 90\% confidence level on the lepton-flavor--averaged branching fraction to be
$${\cal B}(\modepichll)= 2 \times \frac{ \tau_{B^+} }{ \tau_{B^0} } {\cal B}(\modepizll) < 7.9 \times 10^{-8}.$$
We also obtain an upper limit at 90\% confidence level on the lepton-flavor--violating decay \modepiem of
$${\cal B}(\modepiem)< 9.8 \times 10^{-8}.$$

\vfill
\begin{center}

Submitted to the 33$^{\rm rd}$ International Conference on High-Energy Physics, ICHEP 06,\\
26 July---2 August 2006, Moscow, Russia.

\end{center}

\vspace{1.0cm}
\begin{center}
{\em Stanford Linear Accelerator Center, Stanford University, 
Stanford, CA 94309} \\ \vspace{0.1cm}\hrule\vspace{0.1cm}
Work supported in part by Department of Energy contract DE-AC03-76SF00515.
\end{center}

\newpage
} 

\begin{center}
\small

The \babar\ Collaboration,
\bigskip

%
{B.~Aubert,}
{R.~Barate,}
{M.~Bona,}
{D.~Boutigny,}
{F.~Couderc,}
{Y.~Karyotakis,}
{J.~P.~Lees,}
{V.~Poireau,}
{V.~Tisserand,}
{A.~Zghiche}
\inst{Laboratoire de Physique des Particules, IN2P3/CNRS et Universit\'e de Savoie,
 F-74941 Annecy-Le-Vieux, France }
% ...

\end{center}\newpage

\section{Introduction}
\label{sec:Introduction}

\begin{figure}[!hbt]
\begin{center}
\includegraphics[height=5cm]{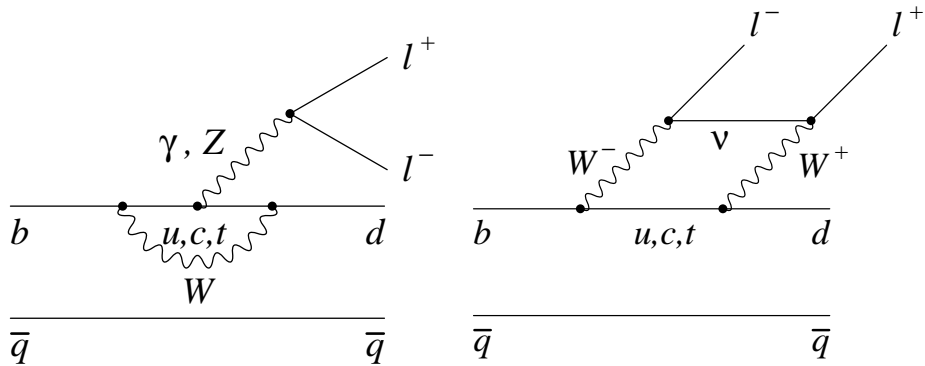}
\caption{Standard Model diagrams for the decays \modepill. For the photon or $Z$ penguin diagrams on the left, boson emission can occur on any 
of the $b$, $t$, $c$, $u$, $d$, or $W$ lines.}
\label{fig:PenguinDiagramsd}
\end{center}
\end{figure}

The decays $B \rightarrow \pi\ell^+\ell^-$, where $\ell^+\ell^-$ is
either an $e^+e^-$ or $\mu^+\mu^-$ pair, are the simplest
manifestations of $b \rightarrow d\ell^+\ell^-$ flavor-changing
neutral currents (FCNC).  In the Standard Model (SM), these decays are
forbidden at tree level and can only occur at greatly suppressed rates
through higher-order processes. At lowest order in the electroweak
couplings, three amplitudes contribute: (1) a photon penguin, (2) a
$Z$ penguin, and (3) a $W^+W^-$ box diagram 
(Fig.~\ref{fig:PenguinDiagramsd}).  Since these decays proceed via
weakly-interacting particles with virtual energies near the
electroweak scale, they provide a promising means to search for
effects from new flavor-changing interactions.  Such effects are
predicted in a wide variety of models, usually in the context of the
similar FCNC $b \rightarrow s\ell^+\ell^-$
~\cite{bib:chargedhiggs,bib:susy,bib:TheoryA,bib:4g,bib:lq}.  
If there
exists non-trivial flavor violation in the new interactions, $b
\rightarrow d\ell^+\ell^-$ can also exhibit large observable effects,
independent of the experimental constraints on $b \rightarrow
s\ell^+\ell^-$~\cite{bib:babarkll2006,bib:sllexp}.

In the Standard Model, the $b\to d$ FCNC decay rate is suppressed
relative to $b \to s$ by a ratio of Cabibbo-Kobayashi-Maskawa matrix
elements $|V_{td}/V_{ts}|^2$, making them especially rare processes
which challenge the sensitivity of the B factories.  The chief
experimental constraint on $\Delta B = 1$ $b \to d$ FCNC 
decays comes from recent
measurements of the exclusive decays $B\to\rho\gamma$ and $B\to
\omega\gamma$~\cite{bib:rhogamma}, which constrain the photon penguin
decay rate to agree with SM predictions with an experimental precision of 
30\%.  The
additional $b\to d$ electroweak amplitudes present in $b \rightarrow
d\ell^+\ell^-$ are weakly constrained experimentally from
older measurements~\cite{bib:mark}.  A recent SM
prediction~\cite{bib:aliev1998} for the \modepichll branching fraction
is based on an operator product expansion combined with $B\to \pi$ 
form factor predictions from light-cone QCD
sum rules; the predicted branching fraction is $3.3\times 10^{-8}$ 
with an uncertainty of about 30\%.  The \modepizll branching
fraction is related to the charged mode by isospin symmetry; accounting
for the difference in charged and neutral $B$ meson lifetimes, the 
branching fraction is given by 
$\frac{1}{2}\times\tau_{B^0}/\tau_{B^+}\times{\cal B}(\modepichll)$.

We also perform a search for the lepton-flavor violating decay 
$B \to \pi e^{\pm} \mu^{\mp}$ which can occur in some models beyond
the Standard Model, such as theories involving leptoquarks~\cite{bib:lq}. 
Earlier searches~\cite{bib:cleo_emu} for these modes have yielded 
no such events. 

\section{The \babar\ Detector and Dataset}
\label{sec:babar}


The data used in this analysis were collected with the \babar\ detector
at the \pep2\ storage ring located at the Stanford
Linear Accelerator Center. 
The data sample comprises 208.9 \invfb\ recorded near the peak of the
$\Upsilon(4S)$ resonance, yielding $(230.1\pm2.5)\times 10^6$ $\BB$ decays, and
an off-resonance sample of 21.5 \invfb\ which is used to study 
continuum background.


The \babar\ detector is described in detail elsewhere~\cite{ref:babar}.
Its most important capabilities for this study
are charged-particle tracking and momentum measurement, 
charged $\pi/K$ separation, and lepton identification.
Charged particle tracking is provided by a five-layer silicon
vertex tracker (SVT) and a 40-layer drift chamber (DCH).
The DIRC, a Cherenkov ring-imaging particle-identification system,
is used to identify charged kaons and pions.
Electrons are identified using the electromagnetic 
calorimeter (EMC), which comprises 6580 thallium-doped CsI
crystals. These systems are mounted inside a 1.5 T solenoidal
superconducting magnet. Muons are identified in the instrumented flux
return (IFR), in which resistive plate chambers are interleaved
with the iron plates of the magnet flux return.

\section{Analysis Method}
\label{sec:Analysis}

\subsection{Event Selection}
We select events that include two oppositely charged leptons
($e^+e^-$, $\mu^+\mu^-$
and $e^{\pm}\mu^{\mp}$) 
as well as a pion (either a $\pi^{\pm}$ 
track or a $\pi^0$ meson decaying to $\gamma \gamma$).
The analysis faces the experimental challenge of isolating a very rare
$B$-meson decay from background due to random combinations of particles
in generic $e^+ e^- \to \FourS \to \BB$ and continuum $e^+ e^- \to q
\bar q$ events ($q \bar q$ being a pair of $u$,$d$,$s$, or $c$ quarks), 
and there is a great abundance of pions and leptons in all of these events.
We use stringent particle identification and a multivariate algorithm
to reduce this background. Another type of background comes from
events that have signal-like features. The main contributions to such
backgrounds are from $B$-mesons decaying to charmonium states
(\textit{e.g.} $\B\to J/\psi\; \pi$ where the $J/\psi$ decays to two
leptons) or $B$-mesons decaying to three hadrons (\textit{e.g.} $B \to D \pi$
with $D \to K \pi$) where two of the hadrons pass the muon
selection. In addition, there is background from the $b \to s
\ell^+\ell^-$ penguin decay $B \to K\; \ell^+
\ell^-$, which has a branching fraction about one order of magnitude
larger than expected for 
$B \to \pi\; \ell^+ \ell^-$~\cite{bib:babarkll2006,bib:aliev1998}. 
These events have a reconstructed $B$ energy lower than expected for signal 
events and are also very efficiently reduced by information from the DIRC
subdetector.

Electrons are required to have a momentum $p > 0.3$ \gevc and are
identified combining information from the
EMC, DIRC, and DCH.
Photons with $E > 30$ \mev that lie within an angular region of 
$50$ mrad in the azimuth angle $\phi$ 
and $35$ mrad in the polar angle $\theta$ 
of the initial electron track direction
are combined with electron candidates
in order to recover energy lost by bremsstrahlung.
We suppress backgrounds due to photon conversions in the $B \to \pi\;
e^+e^-$ channels by removing $e^+e^-$ pairs with invariant mass less
than 30 \mevcc.
Muons with momentum $p > 0.7$ \gevc are identified with a neural
network algorithm using information from the IFR and the EMC.
%

The lepton pairs are combined with a charged or neutral pion candidate
to reconstruct a $B$ meson.  Charged pion candidates are tracks with
specific ionization ($dE/dx$) and Cherenkov angle consistent with a
pion. A likelihood algorithm based on this information reduces fake
rates from kaons down to about $1-5\%$ depending on the track
momentum.  Neutral pion candidates are identified as pairs of
neutral-energy deposits in the EMC, each with an energy greater than
50 MeV in the laboratory frame.  The invariant mass of the pair is
calculated under the assumption that the photons originate from the
$\ell^+\ell^-$ vertex, and is required to satisfy $115 < m_{\gamma
\gamma} < 150$ MeV/$c^2$.


Correctly reconstructed $B$ signal decays produce narrow peaks in the
distributions of two kinematic variables, which can be used to extract
the signal and background yields. For a candidate system of $B$
daughter particles with masses $m_{i}$ and three-momenta ${\bf
p}^*_{i}$ in the $\Upsilon(4S)$ center-of-mass (CM) frame, we define
$m_{\rm ES}= \sqrt{E_{\rm b}^{*2} - |\sum_{i} {\bf p}^*_{i}|^2}$ and
$\Delta E= \sum_{i}\sqrt{m_i^2 + {\bf p}_{i}^{*2}}- E_{\rm b}^*$,
where $E_{\rm b}^*$ is the beam energy in the CM frame.
For signal events, the $m_{\rm ES}$ distribution peaks at the $B$ meson
mass and has a width $\sigma_{m_{\rm ES}} \approx 2.5\ {\rm MeV}/c^2$,
and the $\Delta E$ distribution peaks near zero, with a typical width
$\sigma_{\Delta E} \approx$ 20 MeV.
The mean and width of the $m_{\rm ES}$ 
and  $\Delta E$ distributions 
are determined separately for $e^+e^-$ and $\mu^+\mu^-$ 
modes using data control samples. 
For events reconstructed as 
$e^{\pm}\mu^{\mp}$ we assume the same width and mean as for the $e^+e^-$ modes. 

We define two kinematic regions in terms of $m_{\rm ES}$ and $\Delta E$ 
for signal extraction purposes. 
The signal region is defined to be within $2 \sigma$ of the expected 
mean of the peak in $m_{\rm ES}$ and $\Delta E$. 
The values are given in Table~\ref{tab:signalbox}.
A $2 \sigma$ signal region width was found to be close to optimal 
for both the $S/\sqrt{S+B}$
and $S/\sqrt{B}$ figure of merit, 
where $S$ and $B$ are the number of signal and 
background events expected based on Monte Carlo simulations.

\begin{table}[h]
  \caption{Boundary values defining the signal region for each \modepill mode. 
    The boundaries used in the \modepichem and \modepizem modes are the same 
    as for \modepichee and \modepizee modes, respectively.}
  \label{tab:signalbox}
  \footnotesize
  \begin{center}
    \begin{tabular}{l|rrrr}
      \hline  \hline
      mode         & $m_{\rm ES}$ low & $m_{\rm ES}$ high  & $\DeltaE$ low & $\DeltaE$ high \\
                   & [GeV/$c^2$]      & [GeV/$c^2$]        & [MeV]         & [MeV] \\ \hline
      \modepichee  & 5.2748         & 5.2847           & -53.6        & 37.4\\
      \modepizee   & 5.2767         & 5.2839           & -115.0       & 82.5\\
      \modepichmm  & 5.2749         & 5.2847           & -42.0        & 35.0\\
      \modepizmm   & 5.2764         & 5.2836           & -87.4        & 68.0\\
      \hline  \hline

    \end{tabular}
  \end{center}
\end{table}

We blind ourselves from inspecting the signal region, 
and also a broader region in the $m_{\rm ES}$--$\Delta E$ plane, 
until all selection criteria are optimized and control samples are checked.  
The broader region is used to determine the number of background events from an
unbinned maximum likelihood fit directly from the data.  This fit
region is defined as $5.2 \gevcc < m_{\rm ES}$ and $|\Delta E| < 0.25
\gev$.

Backgrounds arise from three main sources: random combinations of
particles from 
$q\bar q$ and $\tau^+ \tau^-$
events produced in the continuum, random
combinations of particles from $\Upsilon(4S)\to \BB$ decays, and $B$
decays to topologies similar to the signal modes.  The first two,
``combinatorial'', backgrounds typically arise from pairs of
semileptonic decays and produce broad distributions in \mes and
\DeltaE compared to the signal.  The third source arises from modes
which have shapes similar to the signal, such as
$B\to J/\psi \pi$, with $J/\psi\to\ell^+\ell^-$, or $B\to
K\pi\pi$, with kaons or pions misidentified as muons.  
All selection criteria are optimized
with \texttt{GEANT}4~\cite{bib:GEANT} simulated data, independent 
data control samples or with data samples outside the full fit region.


We suppress combinatorial background from continuum processes by selecting 
events with high values in a Fisher discriminant~\cite{bib:Fisher}
constructed as a linear combination of variables 
with coefficients optimized to distinguish between signal
and background. The variables (defined in the CM frame) are
(1) the ratio of second- to zeroth-order Fox-Wolfram
moments~\cite{bib:FoxWolfram} for the event, computed using all
charged tracks and neutral energy clusters;
(2) the absolute value of the angle between the thrust axis of the 
$B$ candidate and that of the remaining particles in the event;
(3) the production angle $\theta_B$ of the $B$ candidate with respect
to the beam axis; and
(4) the ratio of second- to zeroth-order Legendre moments~\cite{bib:Legendre}
for the event, computed using all charged tracks and neutral energy clusters.
These variables exploit the difference between the relatively 
spherical track distribution in $B\bar B$ events and the jet-like 
structure of continuum events. 

We suppress combinatorial backgrounds from $\BB$ events by selecting 
events with high values in a likelihood function constructed
from 
(1) the missing energy of the event, computed from all charged 
tracks and neutral energy clusters; 
(2) the vertex fit probability of all tracks from the $B$ candidate;
(3) the vertex fit probability of the two leptons; and
(4) the angle $\theta_B$ of the $B$ candidate with respect
to the beam axis in the direction of the negative electron beam. 
Missing energy provides the strongest suppression of
combinatorial $B\bar B$ background events, which typically
contain energetic neutrinos from at least two semileptonic $B$ or $D$ 
meson decays.

Both the continuum and $B\bar B$ suppression discriminants are constructed
separately for each $e^+e^-$ and $\mu^+\mu^-$ mode.  
Parameters of the Fisher discriminant have been determined from 
simulated signal events and a data control sample selected below 
the \FourS energy. 
The likelihood function parameters have been determined from
simulated signal events and simulated generic \BB events to 
separate signal events from other types of \BB backgrounds.
For each reconstruction mode, 
the required minimum values of the Fisher and likelihood functions 
are optimized simultaneously by maximizing $S/\sqrt{S+B}$, where 
$S$ is the expected signal yield in the signal region, 
and $B$ is the expected background yield in the signal region 
based on a combination of generic \BB and continuum Monte Carlo 
simulated events.
The background yields are extracted from a two-dimensional, unbinned, 
maximum-likelihood fit to the simulated events in the fit region.
A branching fraction of ${\cal B}(B \to \pi\ell\ell)$ of
$3.3\times10^{-8}$~\cite{bib:aliev1998} is assumed as the signal
estimate.
The optimal selection criteria do not change significantly if we 
alternatively optimize for $S/\sqrt{B}$.
For the $\pi e^{\pm}\mu^{\mp}$ sample we use the same parameters 
and optimal selection points as found for the $e^+e^-$ modes.


The most prominent backgrounds that peak in the signal region 
are $B$ decays to charmonium: $B\to J/\psi\; \pi$ (with $J/\psi
\to\ell^+\ell^-$).  The more abundant $B\to J/\psi\; K$ decay also
contributes as background to $B \to \pi\; \ell^+\ell^-$.
This latter background is strongly suppressed by particle identification
requirements, and such events typically have a reconstructed $B$ energy 
which is lower than expected for our signal mode, 
so that most of these events are outside the signal region for 
$B \to \pi\; \ell^+\ell^-$.
Similarly, $B$ decays to $\psitwos \pi(K)$ final states contribute with a peaking
background component.  We exclude dilepton pairs with dilepton invariant 
mass $m_{\ell^+\ell^-}$ consistent with the
$J/\psi$ mass ($2.90<m_{e^+e^-}<3.20\ {\rm GeV}/c^2$ and
$3.00<m_{\mu^+\mu^-}<3.20\ {\rm GeV}/c^2$) or with the $\psi(2S)$
mass ($3.60<m_{\ell^+\ell^-}<3.75\ {\rm GeV}/c^2$).  This veto is also
applied to $m_{e^+e^-}$ computed without bremsstrahlung photon
recovery.
When a lepton radiates or is mismeasured, $m_{\ell^+\ell^-}$ can shift
away from the charmonium mass, while $\Delta E$ shifts in a correlated
manner. Therefore, we apply an additional veto with $m_{\ell^+\ell^-}$
veto regions defined as a linear function of $\Delta E$.
By extending the veto region this way, we remove nearly all charmonium
events from the fit region and
simplify the shape of the background PDF in the fit region.
For $e^{\pm}\mu^{\mp}$ the charmonium vetoes are defined to be the 
same as for the $e^+e^-$ modes.

\begin{figure}[!hbt]
\begin{center}
  \includegraphics[width=0.9\textwidth]{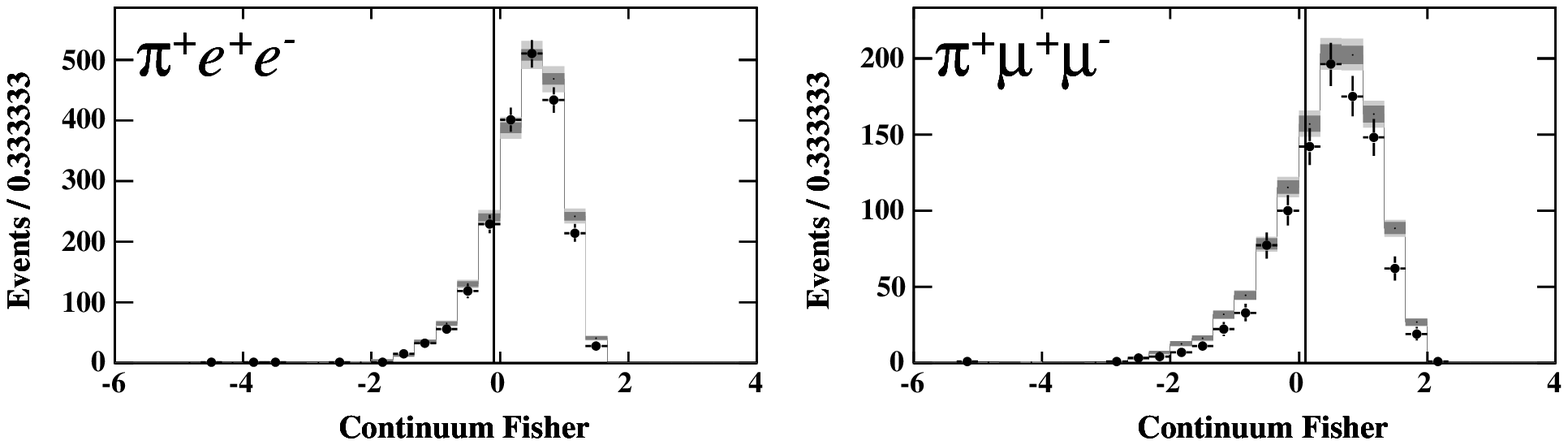}\\
  \includegraphics[width=0.9\textwidth]{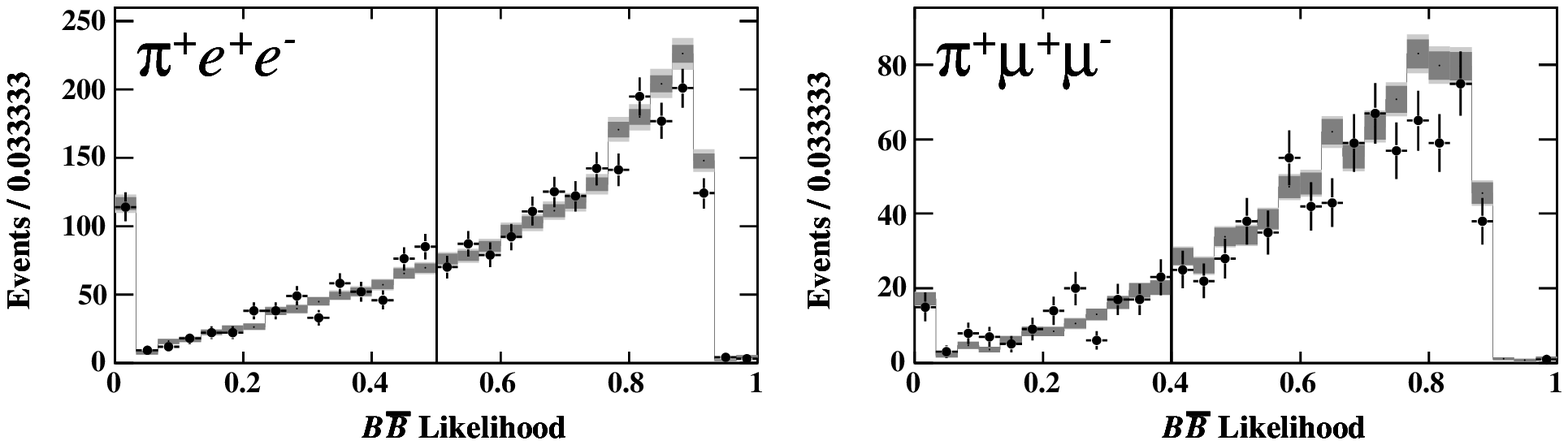}
  \caption{Distribution of the Fisher and likelihood variables in the $J/\psi K^+$ control sample.
  The black points are \babar\ data events and the histograms
  show comparison with simulated events, with statistical and systematic
  uncertainties shown in the dark and light gray bands, respectively.
  The vertical lines indicate the optimal cut values for each variable. 
  Our event selection corresponds to events to the right of the 
vertical line. }
  \label{fig:jpsikcontrol}
\end{center}
\end{figure}

The charmonium events removed by these vetoes are otherwise kinematically 
similar to signal events.  These events therefore serve as 
copious control samples 
for studying signal shapes, selection efficiencies, and
systematic errors. 
Figure~\ref{fig:jpsikcontrol} shows the distributions of the two
background rejection variables for the $B^+ \to J/\psi\; K^+$ control
sample.  This sample is selected with identical requirements as our signal
sample, except the $J/\psi$ veto has been reversed and the hadron
track passes kaon identification. The figure also shows the 
distributions for simulated $B^+ \to J/\psi; K^+$ events and 
the simulations agree well with the data.  
We use this sample to measure the efficiency 
and bound systematic uncertainties of lepton identification, 
Fisher discriminant, and $B$ likelihood requirements. 

In addition to studying the efficiencies using the charmonium control sample,
we measure the branching fractions of ${\cal B}(B \to J/\psi\;\pi)$ 
using \modepill candidates which fail the charmonium veto. 
The measured branching fractions are consistent with the world averages for 
these modes~\cite{PDG}.

In modes with two muons in the final state, 
where the probability for a hadron to be misidentified as a muon can 
be as high as a few percent, background from hadronic
$B$-decays like $B\to D \pi$ with $D\to \pi\pi$ or $D\to
K\pi$ is significant.  These events are suppressed by vetoing
events where the $B \to \pi\; \ell^+\ell^-$ kinematics are consistent
with those of a hadronic $D$ decay. 
We evaluate the $m_{\ell^+\ell^-}$ and $m_{\pi \ell}$ invariant masses 
with $\pi$ or $K$ mass hypotheses on the tracks in all combinations 
corresponding to known $D$-meson decay modes;
we veto those events which have 
two oppositely-charged tracks with an invariant mass within the 
range $1.84 - 1.89$ GeV$/c^2$, 
or events which have a neutral pion and charged track combination with an 
invariant mass within the range $1.79 - 1.94$ GeV$/c^2$.


If more than one charged $B$-meson candidate remain after all
selection requirements have been applied, we select the candidate for
which the pion has the greatest number of hits in the SVT tracking
detector. If more than one neutral $B$-meson candidate remain, we 
select the first one that appears in our data sample ({\textit i.e.} a 
random candidate); the efficiency of this choice is nearly
identical to choosing based on the reconstructed $\pi^0$ mass.  These
criteria are chosen in favor of methods using kinematic information
to avoid any bias due to possible correlations with the signal
extraction variables $m_{\rm ES}$ and $\Delta E$.  More than one $B$
candidate per event occurs in about 13-15\% of charged-$B$ signal
events and 35-60\% of neutral-$B$ signal events, and typically the
multiple candidates are due to more than one possible $\pi$
candidate. For the $e^{\pm}\mu^{\mp}$ modes more than one $B$
candidate occurs in about 5\% of events.


\subsection{Signal Extraction and Background Estimate}
The signal is extracted by counting events in the signal region. 
An upper limit on the branching fraction ${\cal B}(B \to \pi\;\ell^+\ell^-)$ 
is determined from the observed number of events and the expected 
number of background events in this region. We determine the 
background expectation before inspecting the signal region. 


The number of combinatorial background events 
is extracted
by performing a two-dimensional unbinned maximum-likelihood fit to 
$m_{\rm ES}$ and $\Delta E$ in the region $m_{\rm ES}<5.2724\ {\rm GeV}/c^2$
and $|\Delta E|<0.25$ GeV.  
The background probability distribution function (PDF) is modeled 
as the product of 
an ARGUS function~\cite{bib:ARGUS} for $m_{\rm ES}$ and 
an exponential function for $\Delta E$.
The slopes and normalization of these functions are floating in the fit, 
while the endpoint of the ARGUS function is fixed to 5.290 GeV$/c^2$.
The PDF is extrapolated into the signal
region (up to $m_{\rm ES}<5.290\ {\rm GeV}/c^2$) and the expected
number of background events in the signal region is obtained from
integrating the PDF over the signal region. 
The expected number of combinatorial background 
events is given in Table~\ref{tab:systematic bkg}.


We estimate the remaining hadronic $B$ background from a data control
sample of mainly $B\to\pi\pi\pi$ and $B\to K\pi\pi$
events. In the signal sample these events are highly suppressed by the 
lepton identification criteria. 
For the hadronic control sample we select events where the charged tracks 
fail electron and muon identification, except that for data reduction purposes
we require that one track (and only one) pass a looser muon identification 
criteria which still has a high hadron fake rate. 
This selects a sample of predominantly hadronic events. 
These hadronic events are further weighted by their probability 
to pass the signal-selection muon identification, 
by weighting each track with a probability depending on its
particle type hypothesis, momentum and direction.
These track-by-track probabilities are determined from data 
control samples specifically selected to study particle 
identification efficiencies. 
Looking at events with a $\Delta E$ consistent with a correctly
reconstructed $B$-meson, we fit the weighted $m_{\rm ES}$ distribution
for these events to obtain the expected number of hadronic peaking
background events in the signal region.  
We estimate a background of
$0.06\pm0.05$ events in the $\pi^+ \mu^+\mu^-$ channel and
$0.11\pm0.04$ events in the $\pi^0 \mu^+\mu^-$ channel, where the
dominant uncertainty is the statistical uncertainty of the fit.  We
expect no such peaking background in the electron modes or in the
$e\mu$ modes.


$B$-meson decays to $K\, \ell^+ \ell^-$ and $\rho\, \ell^+\ell^-$ 
in the final state are the only other major peaking background 
components expected. The former shift toward lower reconstructed
$B$ energy and are mostly outside the $2 \sigma$ signal region. 
This is the sole largest peaking background component for the
$\pi^+e^+e^-$ mode. The latter contribute even less, since the
reconstructed $B$ in these cases is missing a pion.
These estimates are based on high-statistics samples of simulated 
events. In total, we estimate the number of peaking background events
from leptonic events to range from $0.07\pm0.02$ events for the 
$\pi^+e^+e^-$ mode down to 
$0.02\pm0.01$ events for the $\pi^0 \mu^+ \mu^-$ mode.
No leptonic peaking background is expected in the $e\mu$ modes.

After the vetoes of $B\to J/\psi\, \pi (K)$ and $B\to \psitwos\, \pi (K)$
decays, no remaining background from these are expected, as estimated 
from Monte Carlo simulations. 

Table~\ref{tab:systematic bkg} summarizes the number of background 
events expected in the signal region and the systematic uncertainties 
associated with these estimates. 

\begin{table}[h]
\caption{
  Estimated number of background events and their associated systematic uncertainties. 
  The \mes -\DeltaE fit uncertainties are evaluated by varying the parameters
  of the fit by $\pm 1\sigma$, and the \mes -\DeltaE correlations and \DeltaE shape
  uncertainties are evaluated from using alternative probability density functions. 
  Estimates and systematics from the peaking background are based on Monte Carlo 
  and control sample studies. 
}
\label{tab:systematic bkg}
\footnotesize
\begin{center}
\begin{tabular}{lrrrrrr}
 \hline  \hline
Systematic &
$\pi^+e^+e^-$ & $\pi^0 e^+e^-$ & $\pi^+\mu^+\mu^-$ & $\pi^0 \mu^+\mu^-$ & $\pi^+ e\mu$ & $\pi^0 e\mu$ \\ \hline
\mes -\DeltaE fit  &
$0.89\pm0.31$ & $0.43\pm0.21$ & $0.86\pm0.24$ & $0.22\pm0.18$ & $1.48\pm0.33$ & $1.13\pm0.40$ \\
\mes -\DeltaE correlations  &
$\pm0.02$ & $\pm0.03$ & $\pm0.06$ & $\pm0.03$ & $\pm0.17$ & $\pm0.05$ \\
\DeltaE shape &
$\pm0.03$ & $\pm0.03$ & $\pm0.15$ & $\pm0.02$ & $\pm0.31$ & $\pm0.24$ \\
Leptonic peaking bkg. &
$0.07\pm0.02$ & $0.03\pm0.01$ & $0.04\pm0.01$ & $0.02\pm0.01$ & $0.00\pm0.00$ & $0.00\pm0.00$ \\  
Hadronic peaking bkg. &
$0.00\pm0.00$ & $0.00\pm0.00$ & $0.06\pm0.05$ & $0.11\pm0.04$ & $0.00\pm0.00$ & $0.00\pm0.00$ \\  \hline
Total &
$0.96\pm0.32$ & $0.46\pm0.22$ & $0.96\pm0.30$ & $0.35\pm0.19$ & $1.48\pm0.48$ & $1.13\pm0.47$\\  \hline  \hline
\end{tabular}
\end{center}
\end{table}

\section{Systematic Uncertainties}
\label{sec:Systematics}

In evaluating systematic uncertainties for the branching fractions, we
consider both the uncertainties that affect the signal efficiency estimate, and
uncertainties arising from the background estimate.
Table~\ref{tab:systematic_multi} lists the systematic
uncertainties considered for the signal efficiency.
Sources of uncertainties that affect the efficiency are:
charged-particle tracking efficiency (0.8\% per lepton, 1.4\% per charged
hadron), charged-particle identification (0.7\% per electron pair,
1.9\% per muon pair, 0.5\% per pion), the continuum background and
$\BB$ suppression selection (1.4\%--1.9\% depending on the mode), and
signal simulation statistics (0.1\%).  The estimated number of $\BB$
events in our data sample has an uncertainty of 1.1\%.
An additional systematic uncertainty for the efficiency results from
the choice of form factor model, which alters the $q^2$ distribution
of the signal. We take this uncertainty to be the same as found
in~\cite{bib:babarkll2006}, which uses the same signal model as this
analysis.

Finally, the uncertainties due to the signal efficiency of the \mes and \DeltaE
selection requirements are determined from the measured mean and width of
these distributions in charmonium control samples.  For \modepichll,
we use samples of $B\to J/\psi K^+$ events, in which the mean and width are precisely bounded; for \modepizll, we use 
samples of $B\to J/\psi\pi^0$ events, which have limited statistics and 
introduce a total systematic uncertainty of 7\%.  For the electron modes,
we allow for a larger or smaller bremsstrahlung tail in the \DeltaE 
distribution, introducing a systematic uncertainty of 1-2\%.
The total systematic uncertainty of the signal efficiency is 5\% for 
\modepichll and 9\% for \modepizll. 

\begin{table}[h]
\caption{The sources of systematic uncertainty in signal efficiency (\%) 
considered for $\pi\,\ellell$ decays.}
\label{tab:systematic_multi}
\footnotesize
\begin{center}
\begin{tabular}{lllllll}
 \hline  \hline
Systematic &
$\pi^+e^+e^-$ & $\pi^0 e^+e^-$ & $\pi^+\mu^+\mu^-$ & $\pi^0 \mu^+\mu^-$& $\pi^+e\mu$ & $\pi^0 e\mu$  \\ \hline
Trk eff. &
$\pm3.0$ & $\pm1.6$ & $\pm3.0$ & $\pm1.6$ & $\pm3.0$ & $\pm1.6$ \\
Electron ID &
$\pm0.7$ & $\pm0.7$ &          &          & $\pm0.4$ & $\pm0.4$\\
Muon ID &
         &          & $\pm1.9$ & $\pm1.9$ & $\pm1.0$  & $\pm1.0$\\
Pion ID &
$\pm0.5$ &          & $\pm0.5$ &          & $\pm0.5$ &          \\
$\pi^0$ ID &
         & $\pm3.0$ &          & $\pm3.0$&          & $\pm3.0$ \\
Fisher and $B\bar B$ likelihood &
$\pm1.4$ & $\pm1.4$ & $\pm1.7$ & $\pm1.9$ & $\pm1.4$  & $\pm1.4$\\
MC statistics &
$\pm0.1$ & $\pm0.1$ & $\pm0.1$ & $\pm0.1$ & $\pm0.1$ & $\pm0.1$ \\
$B\bar B$ counting &
$\pm1.1$ & $\pm1.1$ & $\pm1.1$ & $\pm1.1$ & $\pm1.1$ & $\pm1.1$ \\
Model dep.  &
$\pm3.0$ & $\pm3.0$ & $\pm3.0$ & $\pm3.0$ & $\pm3.0$  & $\pm3.0$\\
signal \mes model &
$\pm0.3$ & $\pm5.1$ & $\pm0.4$ & $\pm4.9$ & $\pm0.3$ & $\pm5.1$ \\
signal \DeltaE model &
$\pm0.6$ & $\pm5.1$ & $\pm0.5$ & $\pm5.4$ & $\pm0.5$ & $\pm5.2$ \\
signal \DeltaE radiative tail &
$\pm1.2$ & $\pm1.3$ &  &                  & $\pm1.0$ & $\pm1.4$  \\ \hline
Total &
$\pm4.9$ & $\pm8.8$ & $\pm5.1$ & $\pm9.0$ & $\pm4.9$ & $\pm8.9$ \\  \hline  \hline
\end{tabular}
\end{center}
\end{table}

Systematic uncertainties for the estimated backgrounds arise from two
sources: uncertainties of the maximum likelihood fit which determines
the combinatorial background, and uncertainties of the peaking
background estimates.  Table~\ref{tab:systematic bkg} summarizes the
sources of background systematics.  In the maximum likelihood fit to
the region defined by $5.2 < \mes < 5.2724$\gevcc and $|\DeltaE| <
0.25$\gev, the uncertainty of the extrapolated signal box yield is
determined from the induced change due to $\pm1\sigma$ variations in
the fit region yield, the slope parameter in the \mes PDF, and the
exponent of the \DeltaE PDF.  We also consider the effect of using
different PDF parameterizations on the background estimates, and use
the computed differences to bound a systematic uncertainty.  We fit a
PDF which is correlated in \mes and \DeltaE via a linear \DeltaE
dependence in \mes slope parameter.  We also estimate background from
fits for which the \DeltaE PDF is a linear or quadratic polynomial.
These alternative fits result in differences in estimated background
generally smaller than the uncertainties of the baseline fit. The
uncertainty of the hadronic $B$ decay peaking background is dominated
by the control sample statistics from which it is derived; the
leptonic peaking background uncertainties are dominated by the
uncertainty of the assumed branching fractions for these processes,
particularly \modekll.

\begin{figure}[!hbt]
\begin{center}
\includegraphics[height=0.9\textheight]{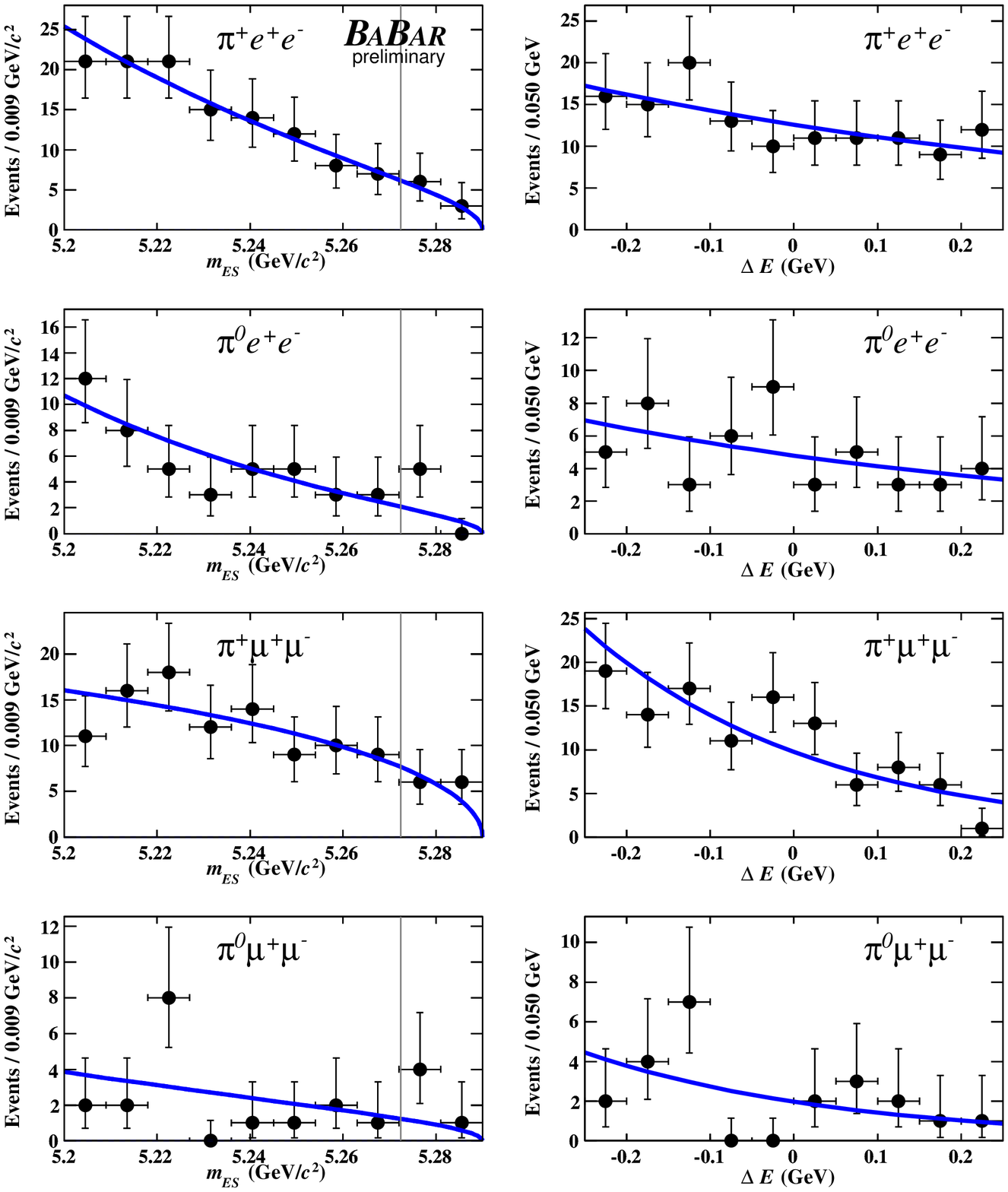}
\caption{
Projections of $m_{\rm ES}$ and $\Delta E$ of \modepill events in the
full fit region. Superimposed is the PDF we use to model combinatorial
background. The parameters were obtained from two-dimensional unbinned
maximum-likelihood fits to events outside the signal region,
with $m_{\rm ES} < 5.2724$ (to the left of the gray vertical line).
} 
\label{fig:2dfits}
\end{center}
\end{figure}

\begin{figure}[!hbt]
  \begin{center}
    \includegraphics[height=0.45\textheight]{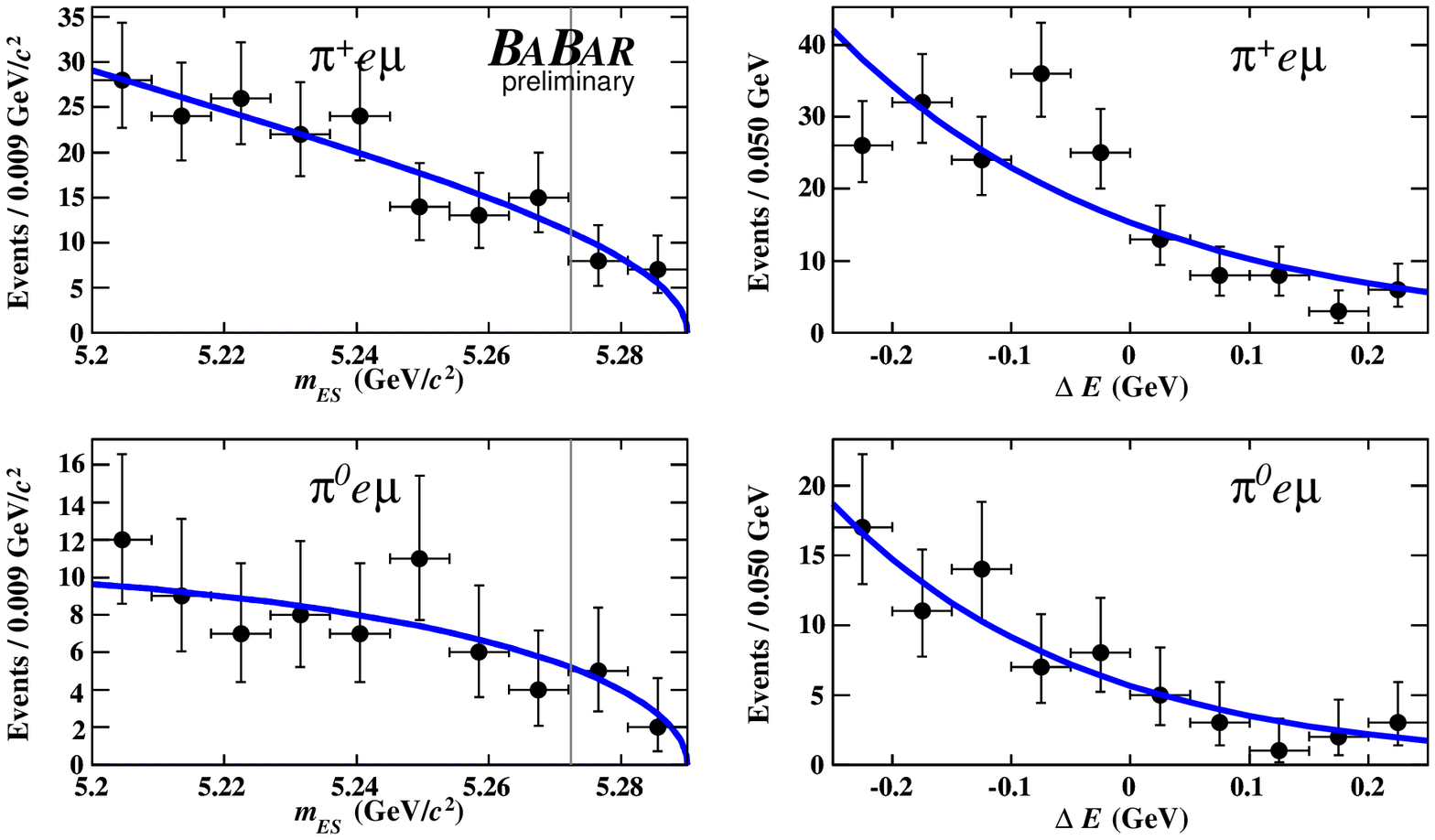}
    \caption{
      Projections of $m_{\rm ES}$ and $\Delta E$ of \modepiem events
      in the full fit region. Superimposed is the PDF we use to model
      combinatorial background. The parameters were obtained from
      two-dimensional unbinned maximum-likelihood fits to events
      outside the signal region with $m_{\rm ES} < 5.2724$ (to the
      left of the gray vertical line).
    }
    \label{fig:2dfitsem}
  \end{center}
\end{figure}

Figures~\ref{fig:2dfits} and~\ref{fig:2dfitsem} show projections 
onto $m_{\rm ES}$ and $\Delta E$ of a two-dimensional unbinned 
maximum-likelihood fit to measure the number of background events. 
The best fit PDFs are consistent with the observed data.  The fit 
region yields for each mode are also consistent with the simulated 
predictions from which optimization of selection criteria were derived.
We estimate the combinatorial background in the signal region from these
PDFs and those estimates together with the peaking backgrounds result
in about one event per decay mode expected in the signal region.

\section{Results}
\label{sec:results}

\begin{figure}[!hbt]
\begin{center}
  \includegraphics[width=\textwidth]{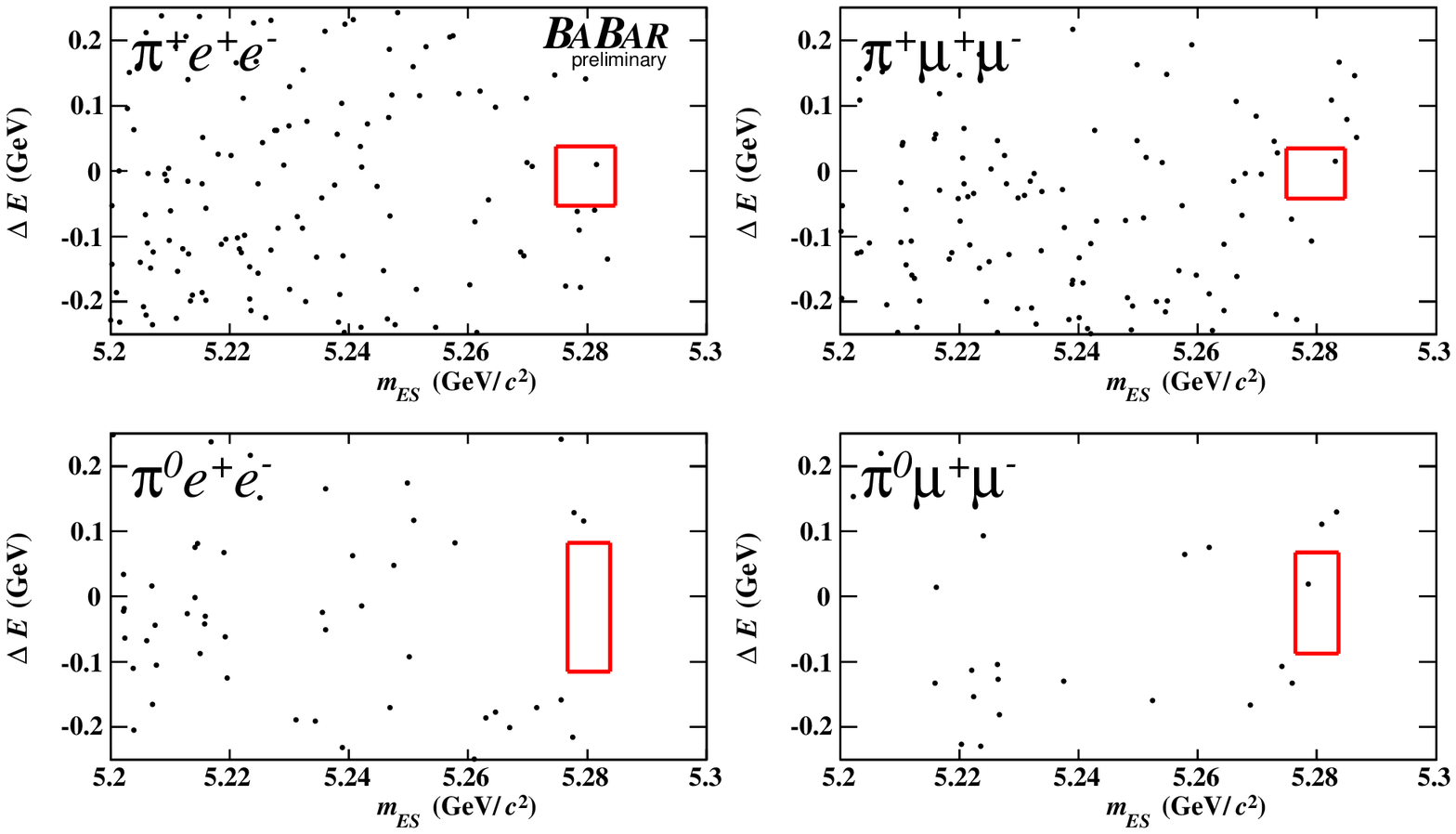}\\
  \includegraphics[width=\textwidth]{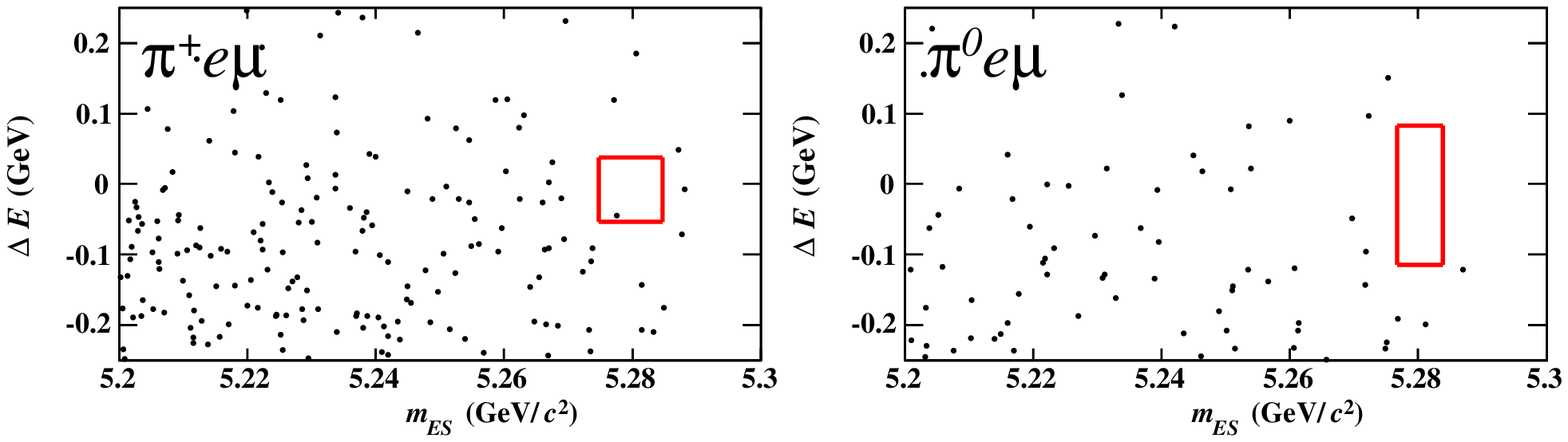}
  \caption{
    Distribution in two dimensions ($m_{\rm ES}$ and $\Delta E$) of events in the background fit region
    passing the selection. The small box defines the signal region from which signal
    candidates are selected.
    \label{fig:scatter}
  }
\end{center}
\end{figure}

Figure~\ref{fig:scatter} shows a scatter plot of the fit region with
events from the \babar\ data that pass all our selection
criteria. The signal events are expected to be found in the signal
region marked by a box drawn in the plot.  We observe a total of 3 \modepill candidates
and 1 \modepiem candidate in the signal region.  This is consistent 
with no significant signal above the expected backgrounds, so we calculate
upper limits at 90\% confidence level via a frequentist method which takes
into account uncertainties both in the signal sensitivity and in the
expected background~\cite{bib:barlow}.  The limits are presented in 
Table~\ref{tab:results}, along with all numbers necessary in this method to 
perform the calculation.  The upper limits for the electron and electron-muon 
modes are in the range of $1-2\times10^{-7}$; the muon modes are somewhat less
sensitive with limits in the range of $2-5\times10^{-7}$.
Assuming the partial widths of \modepill to electrons and muons are equal, 
the limits for the two decay modes can be simply combined to provide a 
combined limit of $1.06\times 10^{-7}$ for \modepichll and 
$1.02\times 10^{-7}$ for \modepizll.  Assuming further that there is 
isospin asymmetry in the partial widths of \modepill to charged and neutral 
pions, the four \modepill modes can be combined with the constraint 
${\cal B}(\modepizll) = \frac{1}{2}\times\tau_{B^0}/\tau_{B^+}\times{\cal B}(\modepichll)$ 
to compute a combined limit, expressed in terms of the \modepichll
branching fraction, of $0.79\times10^{-7}$.  A similar combined 
limit of $0.98\times 10^{-7}$ is obtained for the lepton-flavor 
violating mode \modepiem. 

Table~\ref{tab:results} also lists the mean expected upper limits, 
defined as a sum of upper limits obtained for $n$ events observed 
($n = 0,1,2...$), each weighted by the Poisson probability of observing 
$n$ events when expecting background only. 

\begin{table}[tbp!]
 \caption{ Results for the \modepill analysis, including observed 
signal candidate events,  expected background, signal yield
upper limit at 90\% confidence level, signal efficiency, 
expected branching fraction upper
limit at 90\% confidence level, 
and the observed branching fraction upper limit at 90\% confidence level.
The numbers in parentheses are limits evaluated without the inclusion of 
systematic uncertainties. 
Combined limits at the bottom are derived from simultaneous limits calculated
from the individual modes.}
 \label{tab:results}
 \footnotesize
 \begin{center}
 \begin{tabular}{lcccccc}
\hline\hline
 &\multicolumn{1}{c}{Observed} & \multicolumn{1}{c}{Expected}
  & \multicolumn{1}{c}{Events UL}
  & \multicolumn{1}{c}{Signal}
  & \multicolumn{1}{c}{Expected BF UL}
  & \multicolumn{1}{c}{BF UL}\\
\multicolumn{1}{c}{Mode}   & \multicolumn{1}{c}{Events} & 
   \multicolumn{1}{c}{Background} &
  \multicolumn{1}{c}{90\% C.L.} &
  \multicolumn{1}{c}{Efficiency} &
  \multicolumn{1}{c}{90\% C.L. ($10^{-7}$)}
  & \multicolumn{1}{c}{90\% C.L. ($10^{-7}$)}\\
  \hline 
  \modepichee         &$1$&  $0.96\pm0.32$& $2.97$ ($2.93$) &$7.5\pm0.4\%$&1.64& $1.72$ ($1.70$)  \\
  \modepizee          &$0$&  $0.46\pm0.22$& $1.86$ ($1.84$) &$6.3\pm0.6\%$&1.79& $1.29$ ($1.27$)  \\
  \modepichmm         &$1$&  $0.96\pm0.30$& $2.96$ ($2.93$) &$5.2\pm0.3\%$&2.37& $2.47$ ($2.45$)  \\
  \modepizmm          &$1$&  $0.35\pm0.19$& $3.57$ ($3.55$) &$3.4\pm0.3\%$&3.18& $4.56$ ($4.53$)  \\
  \modepichem         &$1$&  $1.48\pm0.48$& $2.49$ ($2.41$) &$6.3\pm0.3\%$&2.17& $1.72$ ($1.66$)  \\
  \modepizem          &$0$&  $1.13\pm0.47$& $1.28$ ($1.18$) &$3.7\pm0.3\%$&3.52& $1.50$ ($1.38$)  \\
 \hline
  \modepichll        &&&& && $1.06$ ($1.04$)\\
  \modepizll         &&&& && $1.02$ ($1.01$)\\
  \modepill          &&&& && $0.79$ ($0.77$)\\
  \modepiem          &&&& && $0.98$ ($0.90$)\\
\hline\hline
 \end{tabular}
 \end{center}
\end{table}

\section{Conclusion}
\label{sec:conclusion}
We present a preliminary result of a search for $B \to \pi
\ell^+\ell^-$ using a sample of $(230.1\pm2.5)\times 10^6$ $\BB$ pairs
produced at the \FourS resonance. We do not see any excess of events
in the signal region, and we measure the upper limit at 90\%
confidence level of the lepton-flavor--averaged branching fraction to
be
$${\cal B}(\modepichll)= 2 \times \frac{ \tau_{B^+} }{ \tau_{B^0} } {\cal B}(\modepizll) < 7.9 \times 10^{-8},$$
consistent with the Standard Model prediction of $3.3\times 10^{-8}$.
This is the first such search performed by the B factory experiments; 
with anticipated final samples of order 1 ab$^{-1}$, and with the small 
backgrounds observed in this analysis, it may be possible in the future to
achieve an experimental sensitivity comparable to the Standard Model 
prediction.

\section{Acknowledgments}
\label{sec:acknowledgments}
We are grateful for the 
extraordinary contributions of our \pep2\ colleagues in
achieving the excellent luminosity and machine conditions
that have made this work possible.
The success of this project also relies critically on the 
expertise and dedication of the computing organizations that 
support \babar.
The collaborating institutions wish to thank 
SLAC for its support and the kind hospitality extended to them. 
This work is supported by the
US Department of Energy
and National Science Foundation, the
Natural Sciences and Engineering Research Council (Canada),
Institute of High Energy Physics (China), the
Commissariat \`a l'Energie Atomique and
Institut National de Physique Nucl\'eaire et de Physique des Particules
(France), the
Bundesministerium f\"ur Bildung und Forschung and
Deutsche Forschungsgemeinschaft
(Germany), the
Istituto Nazionale di Fisica Nucleare (Italy),
the Foundation for Fundamental Research on Matter (The Netherlands),
the Research Council of Norway, the
Ministry of Science and Technology of the Russian Federation, and the
Particle Physics and Astronomy Research Council (United Kingdom). 
Individuals have received support from 
the Marie-Curie IEF program (European Union) and
the A. P. Sloan Foundation.


\begin{thebibliography}{99}
\bibitem{bib:Legendre}{The $i$-th moment $L_i$ is defined by
$L_i = \sum_j |p^*_j| |\cos(\theta^*_j)|^i$,
where the $p^*_j$ are the CM momenta of all particles not
used in reconstructing the signal $B$ candidate, and the angle
$\theta^*_j$ is
between the particle's momentum and the thrust axis of the signal
$B$. }
\end{thebibliography}
\end{document}